\documentclass[aps,prl,twocolumn,showpacs,groupedaddress,superscriptaddress,longbibliography]{revtex4-1}

\usepackage{natbib,hyperref}
\hypersetup{colorlinks=true, citecolor=blue, urlcolor=blue, linkcolor=blue}
\usepackage[english]{babel}
\usepackage[utf8]{inputenc}
\usepackage{graphicx}
\usepackage[caption=false]{subfig}
\usepackage{amsmath}
\usepackage{amsfonts}
\usepackage{amssymb}
\begin{document}

\title{Simultaneous control of multi-species particle transport and segregation in driven lattices}

\author{Aritra K. Mukhopadhyay}
 \email{Aritra.Mukhopadhyay@physnet.uni-hamburg.de}
 \affiliation{Zentrum f\"ur Optische Quantentechnologien, Universit\"at Hamburg, Luruper Chaussee 149, 22761 Hamburg, Germany}
 \author{Benno Liebchen}
  \email{liebchen@hhu.de}
\affiliation{SUPA, School of Physics and Astronomy, University of 
Edinburgh, Peter Guthrie Tait Road, Edinburgh, EH9 3FD, UK}
 \affiliation{Institute for Theoretical Physics II: Soft Matter, Heinrich-Heine University D\"usseldorf, Universit\"atsstrasse 1, 40225, D\"usseldorf, Germany} 
 \author{Peter Schmelcher}
  \email{Peter.Schmelcher@physnet.uni-hamburg.de}
 \affiliation{Zentrum f\"ur Optische Quantentechnologien, Universit\"at Hamburg, Luruper Chaussee 149, 22761 Hamburg, Germany}
 \affiliation{The Hamburg Centre for Ultrafast Imaging, Universit\"at Hamburg, Luruper Chaussee 149, 22761 Hamburg, Germany}

\date{\today}

\begin{abstract}
We provide a generic scheme to separate the particles of a mixture by their physical properties like mass, 
friction or size. The scheme employs a periodically shaken two dimensional dissipative lattice and hinges on a simultaneous 
transport of particles in species-specific directions. 
This selective transport is achieved by controlling the late-time nonlinear particle dynamics, 
via the attractors embedded in the phase space and their bifurcations.
To illustrate the spectrum of possible applications of the scheme, we exemplarily 
demonstrate the separation of polydisperse colloids and mixtures of cold thermal alkali atoms in optical lattices. 
\end{abstract}


\maketitle

\paragraph{Introduction} 
The controlled separation, or spatial sorting, of particle mixtures based on their physical properties like mass, size, shape or mobility presents 
major challenges cutting across disciplines from biomedical problems such as the 
separating of malignant circulating tumour cells from leucocytes in the bloodstream \cite{Jin2014}
to technological problems on colloidal and granular scales \cite{Hanggi2009}. Following these challenges, much effort has been devised to develop innovative separation schemes
complementing traditional techniques such as filtration, distillation or evaporation of mixtures. 
For example, to separate heterogeneous granular particle mixtures
in geological and biological systems, it has been shown that vibrating a substrate \cite{Mullin2000, Shinbrot1998, Hong2001} allows to 
separate two species by size. 
To separate particles on colloidal scales which are significantly affected by Brownian noise, it has been shown that suitable 
external forcing may be sufficient to separate mixtures \cite{Bouzat2010, Zeng2010, Hanggi2009, Marchesoni1998}.

\begin{figure*}[t]
\includegraphics[scale=0.035]{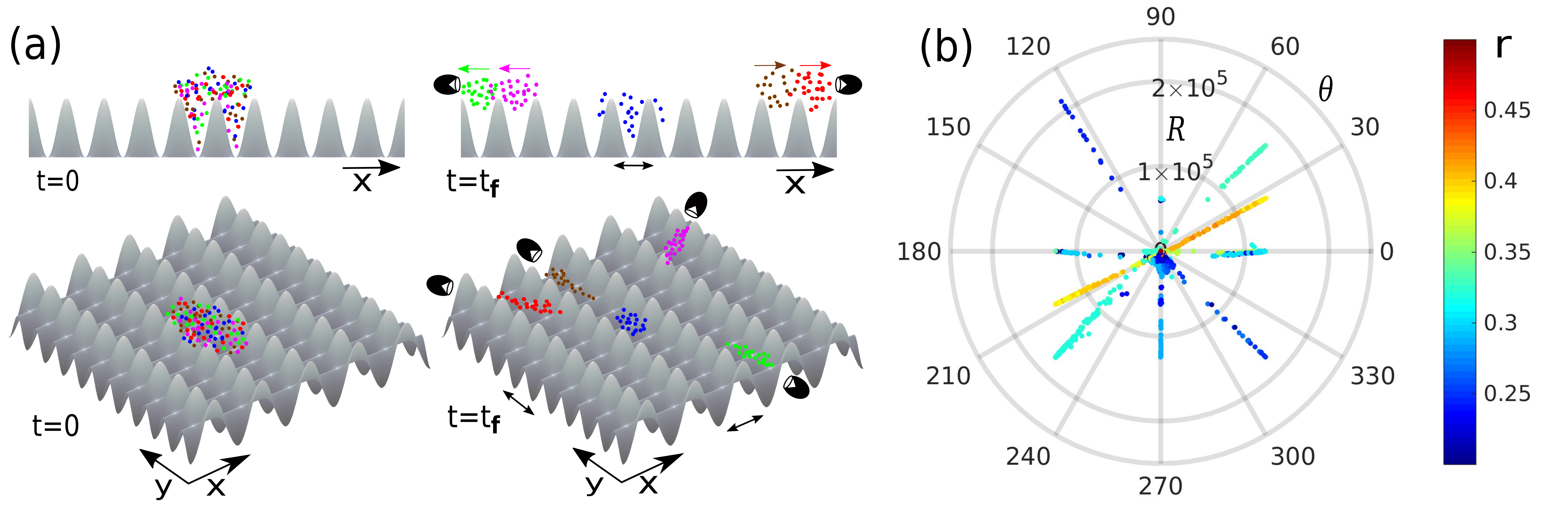} 
\caption{(a) Upper panel: Cartoon of a standard particle segregation scheme based on 1D ratchets allowing for the separation 
of a mixture of particles (left) with different
mass, mobility or radius (color) into two components (right figure).  
Lower panel: The present scheme allows for a controlled angular-specific transport of many species which can be applied
to separate or sort multi-species mixtures. (b) Snapshot of the positions  (in the radial $R$ and angular $\theta$ coordinates) of the particles in the mixture with different radii $r$ (shown in colorbar) at $t=5\times10^4 T$. Particles in at least four different radii ($r$) intervals can be separated at different angles ($\theta$; position of detector) using this setup: the pairs of the radii intervals and angles are ($0.30\lesssim r\lesssim 0.32$, $\theta \approx 45^{\circ}, 225^{\circ}$), ($0.38\lesssim r\lesssim 0.40$, $\theta \approx 30^{\circ}, 210^{\circ}$), ($0.25 \lesssim r \lesssim 0.26$, $\theta \approx 120^{\circ}$) and  ($0.28 \lesssim r \lesssim 0.29$, $\theta \approx 270^{\circ}$). The remaining parameters are $V=0.17, \eta=0.01$, $d_x=d_y=1$, $\omega=1$ and $2k_x=k_y=2$. The other particles are confined at much smaller distances near the origin and one can repeat the process with a different value of $V$ in order to segregate them (to be discussed later). }\label{fig1}
\end{figure*}

One important class of innovative separation schemes employs so-called Brownian ratchets in 
which thermal Brownian motion combined with external time-dependent 
driving generates directed particle motion \cite{Astumian2002, Hanggi2009, Hanggi2005}.
Based on an appropriate design of such ratchets, the direction of the emerging particle current may depend on `internal' particle properties 
such as mass, radius or mobility. This dependence can be exploited to simultaneously transport two different particle species in opposite direction, 
i.e. for separating them (Fig.~\ref{fig1}a, upper panel).
Based on this idea, it has been possible to establish a rich set of schemes to separate two component particle mixtures, including
a massively parallel particle filter \cite{Matthias2003} serving as an artificial microsieve
with potential biomedical applications \cite{Hanggi2009}
and schemes allowing to separate mixtures of cellular membrane associated molecules that differ in electrophoretic 
mobility and diffusion coefficient \cite{VanOudenaarden1999}.
Further examples of ratchet based separation devices allow for the size-sorting of 
superparamagnetic particles using periodically switching magnetic fields \cite{Liu2016}, 
of active and passive particles using active ratchet systems \cite{Reichhardt2017, Ai2016} and heterogeneous cell mixtures using 
microfluidic funnel ratchets \cite{McFaul2012}. 
Finally, we note that ratchets in superlattices also allow to separate particles by the type of their motion (ballistic/chaotic) and allow to 
sort (or filter) them by velocity \cite{Petri2011a, Wulf2012}.

These examples illustrate the rich versatility of 
ratchet-based separation schemes which, in the following respects, seems to be somewhat advantageous over traditional techniques like sieving or filtrating:
(i) they may operate on many technologically relevant scales ranging from 
granular and colloidal sizes down to the nanoregime \cite{Hanggi2009} and even to the size of single atoms \cite{Renzoni2009},
(ii) the particle current underlying particle separation can be controlled with an external field \cite{Reimann2002}, even in real time \cite{Liebchen2012, Mukhopadhyay2016}, and (iii) they allow to separate particles with respect to all kinds of physical properties from size and mass to charge and mobility.
However these advantageous, ratchet-based schemes seem to have one striking disadvantage: 
unlike sieves which can be easily stacked to sort many-component mixtures by size, 
most (if not all) ratchet-based schemes are restricted to the separation of only two species, based on forward transport of 
one species and reverse transport of the other one. Since combinations of several ratchets, each separating two species, would be rather sophisticated to design and produce, it would be desirable to know a mechanism allowing for a simultaneous separation of many species.

Here we propose such a mechanism unifying the above advantages of ratchet-based separation schemes with the ability to efficiently separate multi-species mixtures (Fig.~\ref{fig1}a) based on their physical properties like mass, friction or size. To develop this scheme, we use a periodically shaken two dimensional dissipative lattice, which can be produced for example based on optical molasses and counterpropagating laser beams using oscillating mirrors or acousto-optic modulators (AOMs) \cite{Renzoni2009}, 
and establish a new route to simultaneously control the transport angle of each species in a mixture individually. The new idea is to exploit the strongly nonlinear character of driven lattices to control the late-time particle dynamics species-selectively on the phase space level, which would be difficult or impossible when using standard overdamped frameworks. Initializing particles representing mixtures, of say, polydisperse colloids differing in the friction they experience, we demonstrate that each colloidal species travels in an individual direction through the lattice allowing for their collection with an angular detector (or a reservoir). The segregation scheme should apply more generally to particles on atomic and nano scales up to colloids and granular particles. In particular, contrasting many other ratchet-based separation schemes, the present one does not hinge on Brownian noise (but is robust against it) and should hence apply for granular particles which are too large to experience a significant effect from Brownian noise and in principle also for the extreme case of a mixture of cold thermal atoms with different masses in purely optical setups \cite{Renzoni2009}, where a noise-providing medium is practically absent.

\paragraph{Setup} 
We consider a mixture of $N$ non-interacting classical particles in a two 
dimensional lattice defined by a periodic potential $V(x,y)=V\cos k_x x (1+\cos k_y y)$ which is driven via 
external bi-harmonic driving forces $f_{x,y}(t)= d_{x,y} (\cos \omega t + 0.25\cos (2\omega t + \pi/2))$ acting in both $x$ and $y$
directions and breaking parity $\bf x\rightarrow -\bf x + \pmb{\chi}$ and time-reversal $t\rightarrow -t +\tau$ symmetry, with additional constant spatial and temporal shifts, to allow for a directed particle transport \cite{Reimann2002,Flach2000}. Here, $d_x, d_y$ denote the respective driving amplitudes in the two directions, 
$k_x, k_y$ the respective wave numbers and $\omega$ is the frequency of the external driving force. The system thus has spatial and temporal 
periodicities of $L_{x,y}=2\pi / k_{x,y}$ and $T=2\pi / \omega$.  Such a two dimensional lattice potential can be created for example in cold atom setups by using two sets of counter propagating laser beams of non-orthogonal polarizations between mirrors and the driving can be implemented with standard techniques 
like acousto-optical modulators and radio frequency generators leading to a lateral oscillation of the mirrors and hence of the lattice \cite{Struck2013}. Here, the damping can be realized using optical molasses \cite{Grynberg2001}.

Introducing dimensionless variables $x'=k_x x$, $y'=k_y y$ and $t'=\omega t$ and dropping the primes for simplicity, the equation of motion for a single particle of mass $m$ located at position $\mathbf x$ with momentum $\bf p$ in such a setup reads 
\begin{eqnarray}
\ddot{\bf x} = &\ & U_{x}\sin x(1+\cos y) {\bf e}_x + U_{y}\cos x \sin y {\bf e}_y \nonumber \\
			   &+&   (\cos t + 0.25 \cos (2 t + \pi/2)) {\bf F} - \Gamma \dot{\bf x} + \pmb{\xi} (t)
\end{eqnarray}
where ${\bf e}_x=(1,0)$ and ${\bf e}_y=(0,1)$. The parameter space of this model has five essential dimensions with
$U_{x,y}=\frac{Vk_{x,y}^2}{m\omega ^2}$ comparing the velocity of a particle in a static lattice with the 
velocity of the oscillating lattice, {\bf F}=$\left( \frac{k_x d_{x}}{m\omega ^2},\frac{k_y d_{y}}{m\omega ^2} \right) $ being a reduced driving amplitude and 
$\Gamma = \frac{\gamma}{m\omega}$ comparing the relaxation time due to dissipation (in the underlying static lattice) with the timescale of the lattice oscillation.
$\pmb{\xi} (t)=(\xi_x,\xi_y)$ denotes thermal fluctuations modelled by Gaussian white noise of zero mean with the property $\langle \xi_\alpha (t) \xi_\beta (t')\rangle = 2 D\delta_{\alpha \beta}\delta (t-t')$ where $\alpha,\beta \in {x,y}$ and $D$ is the dimensionless noise strength. In the following, we are mainly interested in the underdamped regime where dissipation is weak but important and entirely neglect Brownian noise in most of our simulations assuming low temperatures in the case of cold atoms or large particle masses as e.g. for granular particles or large underdamped colloids. Our main results are all robust against typical noise as we detail further below. 

\paragraph{Simultaneous control of directed transport and particle segregation}

Before detailing the general working principle of the scheme, we first illustrate its application to a polydisperse mixture of $N=2\times10^4$ underdamped colloidal particles with continuously and uniformly distributed random radii $r \in [0.2,0.5]$. We assume that the only effect that the radii has is that it governs both the colloidal mass ($m=\frac{4}{3}\pi \rho r^3$) and the dissipation coefficient ($\gamma=6\pi \eta r$) which the colloids experience in the surrounding solvent. While $r$ may generally also affect the interaction between the colloids and the lattice potential, we treat them as point like for our computation since for e.g., a $CO_2$ laser, the lattice spacing would be $10^1-10^3$ times larger than typical colloidal sizes. 

\begin{figure}
\includegraphics[scale=0.08]{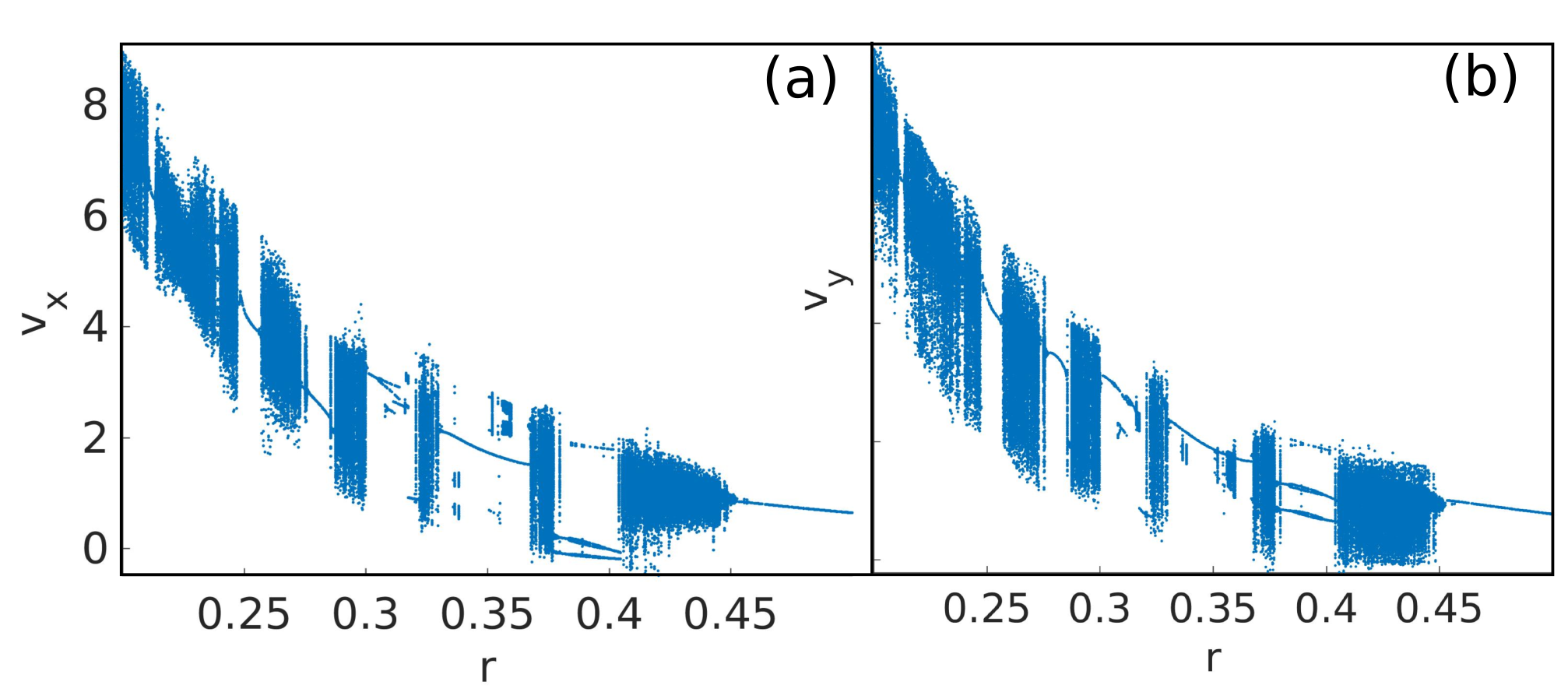}
\caption{Bifurcation diagrams of the velocity of particles for varying radii initialized at the potential minima $x=y=\pi$ with $\mathbf{v}=0$. The regions with broad distributions of velocities (e.g. the radii in the range $0.41\lesssim r\lesssim 0.45$) denote chaotic attractors. The intervals of radii corresponding to a  single velocity value (see $r\approx 0.275$) denote period 1 limit cycles whereas those corresponding to several values of the velocity denote multi-periodic limit cycles (see e.g., particles with $r\approx 0.39$ ending up in a period 2 limit cycle). Remaining parameters are the same as in Fig.~\ref{fig1}b.}\label{fig3}
\end{figure}

We initialize the particles randomly within a square region $L_x \times L_y$ of the lattice and give them small random velocities. 
To mimic potential experimental scenarios, we first allow the colloids to equilibrate in a static lattice for a time  $1000T$ (Fig.~\ref{fig1}a, lower panel; left).
Now switching on the driving, and waiting until  
$\sim 5\times10^4 T$ we observe 
ballistic particle jets (`rays'), radially moving away from their initial positions at different angles (Fig.~\ref{fig1}b). Strikingly, most of 
these `rays' have an almost uniform color (Fig.~\ref{fig1}b), meaning that they involve only colloids with very similar radii and separate them from the rest of the mixture. The figure shows the simultaneous separation of four `species' by radius, while the rays at 0 and 180 degrees, as well as the particle cloud around the origin, still contain a mixture of particles with different radii which may be considered as `losses'. When analyzing the general working principle of our scheme below, it will become clear that the presented segregation is not at all limited to the four specific radii intervals separated so far, but that it is tunable and can be applied to separate particles with a desired set of radii from the mixtures.

%



\paragraph{Discussion}
We now analyze the working principle underlying the observed particle separation. 
The separation principle can be best understood on the level of the invariant manifolds in the five dimensional phase space (${x,y,p_x,p_y,t}$) of the system.
In dissipative systems like the one described here, the asymptotic $t\rightarrow \infty$ dynamics
depends on the set of attractors (stable invariant manifolds) in the phase space and the associated
sets of all initial conditions which asymptotically end up in an attractor i.e the basin of attraction.
For the present periodically driven system, there are two relevant classes of attractors:
\textit{chaotic attractors} and \textit{limit cycles}, representing chaotic and periodic motion respectively. Generally, the set of attractors
and their basins of attraction depend on the parameters of the system, including particle-related parameters such as their mass.

The idea here is to tune the set of attractors such that the late-time dynamics of particles with different properties 
is governed by species-dependent limit cycles which transport particle to species-specific directions. 
Generally, limit cycle attractors transport particles in a well-defined direction and with a characteristic 
(quasi)periodic velocity with average $\mathbf{\bar{v}}\equiv (\bar{v}_x,\bar{v}_y) =\left(\frac{n_x L_x}{m_x T} , \frac{n_y L_y}{m_y T}\right) $ 
where $n_x,m_x,n_y,m_y$ are attractor specific integers. 
Hence at a large distance from the centre of the square region $L_x \times L_y$ where the particles were initialized (which we henceforth refer to as `origin'), particles following the dynamics of a limit cycle attractor can be collected by placing a 
suitable collector (detector or reservoir) at an angle $\theta= \tan^{-1} \frac{\bar{v}_y}{\bar{v}_x}=\tan^{-1} \frac{m_x n_y L_y}{m_y n_x L_x}$. Thus, if we manage to tune the limit cycle attractors and their basins of attraction in the phase space such that particles with e.g. different radii 
end up in different limit cycle attractors they will be automatically separated (Fig.~\ref{fig1}b). The size of the detectors would depend on the angular spreading $\bigtriangleup \theta$ of the particle `jets', which for this setup (Fig.~\ref{fig1}b) was found to be of the order of $0.01$ degrees. Hence, if one places the detectors at a radial distance of $R=10^5$ from the origin, their required sizes would be determined by the arc length $R \bigtriangleup \theta \sim 4L_x$. Such detection and tracking of colloidal particles are routinely done using high resolution optical tracking and holographic microscopy \cite{Garbow1997, Lee2007}. The chaotic particles, owing to their diffusive nature, stay much closer to the origin and as a 
result do not interfere with the segregation process.

In order to predict whether a particle would end up in a chaotic attractor or a limit cycle we compute the 
`bifurcation diagram' associated with the particle velocity as a function of its radius. To do this, we initialize particles of
different radii and after an initial transient time stroboscopically monitor (at multiples of $T$) the velocity as a function of the radius. 
The resulting bifurcation diagrams (Fig.~\ref{fig3}) shows that particles can be either attracted to chaotic attractors, period 1 limit cycles or 
multi-periodic limit cycles. Hence for a given set of parameters, one can predict from such bifurcation diagrams the angles to which particles of 
different radii would travel. 

\begin{figure}
\includegraphics[scale=0.1]{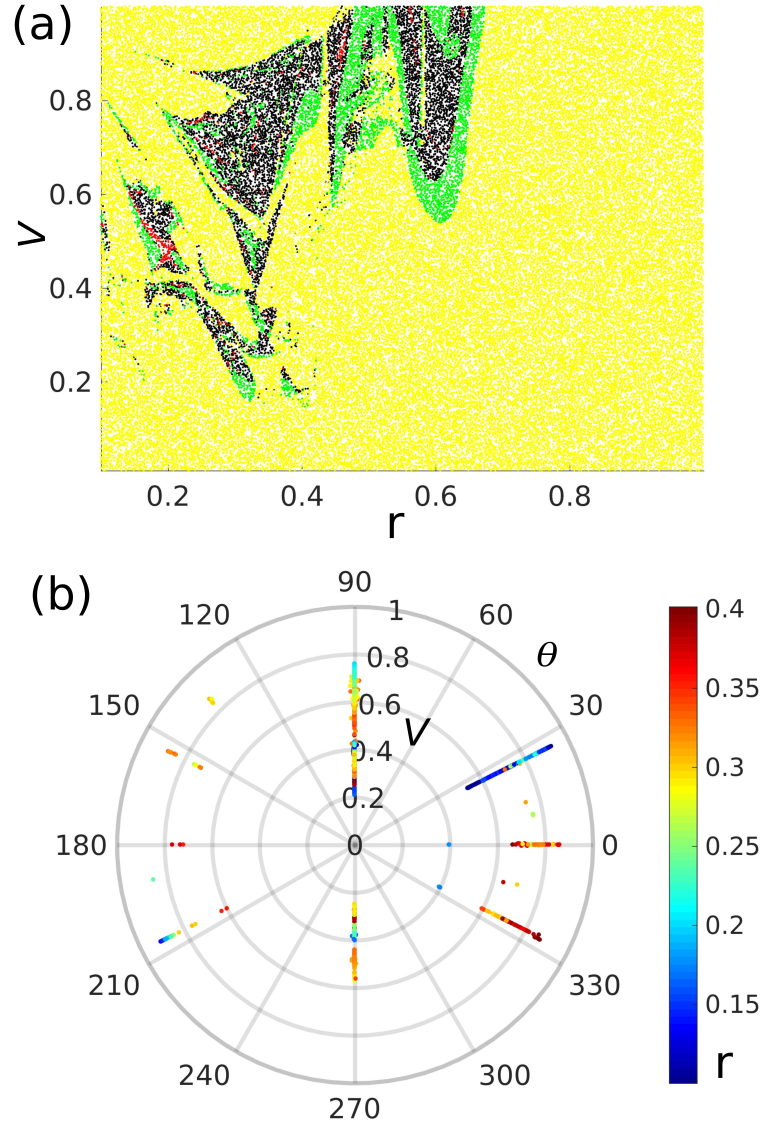}
\caption{(a) Bifurcation diagram of $v_\theta$ (in colour) as a function of lattice potential height $V$ and particle radius $r$. The yellow regions denote period 1 limit cycles, green: period 2 limit cycles, red: period 3 limit cycles and black denotes more than three limit cycle attractors. (b) Asymptotic direction of flight $\theta$ (in the angular coordinates) of the particles ending up in limit cycle attractors (corresponding to yellow, green and red regions of Fig.~\ref{fig4}a) with average velocity $\mathbf{\bar{v}}\neq 0$ for different values of $V$ (in the radial coordinates) and $r$ (in colorbar). Hence, one can predict a priori that  a lattice with a potential height of e.g. $V=0.6$ can separate particles with radii in the intervals $0.26 \lesssim r \lesssim 0.3$ to $\theta \approx 90^{\circ}$ and analogously $0.12 \lesssim r \lesssim 0.16$ to $\theta \approx 27^{\circ}$ and $0.36 \lesssim r \lesssim 0.33$ to $\theta \approx 227^{\circ}, 333^{\circ}$. Remaining parameters are $\eta=0.05$, $d_x=d_y=1$, $\omega=1$ and $2k_x=k_y=2$.} \label{fig4}
\end{figure}

To make the segregation scheme more flexible and separate particles of a varying range of sizes we use the lattice potential height $V$ as a control parameter. Using similar concept as above, one can construct a two parameter bifurcation diagram of the particle velocity $v_\theta=\tan^{-1} \frac{v_y}{v_x}$ (in the angular coordinates) showing whether a particle of radius $r$ ends up in a limit cycle attractor or a chaotic attractor for a given lattice potential height $V$ (Fig.~\ref{fig4}a). For our segregation scheme, we choose those values of $V$ and $r$ for which the particle dynamics is asymptotically governed by the limit cycle attractors. Depending on the average velocity $\mathbf{\bar{v}}$ of these limit cycles, particles of a given radius might be trapped ($\mathbf{\bar{v}}=0$) in the lattice or fly out ballistically at certain angles ($\mathbf{\bar{v}}\neq 0$). The asymptotic direction of flight $\theta= \tan^{-1} \frac{\bar{v}_y}{\bar{v}_x}$ for these ballistic particles as a function of $V$ and $r$ (Fig.~\ref{fig4}b) predicts a priori at which angle a particle of a given radius would travel for a chosen lattice height $V$. At this stage, the segregation protocol becomes very simple: given a set of particles with different radii, the task is to simply choose a value of $V$ from Fig.~\ref{fig4}b for which the different species travel ballistically at different angles. One can in principle repeat this process with a different value of $V$ in order to also separate the cloud of `lost' particles, which owing to their diffusive motion, remain close to the origin.


\begin{figure}
\includegraphics[scale=0.13]{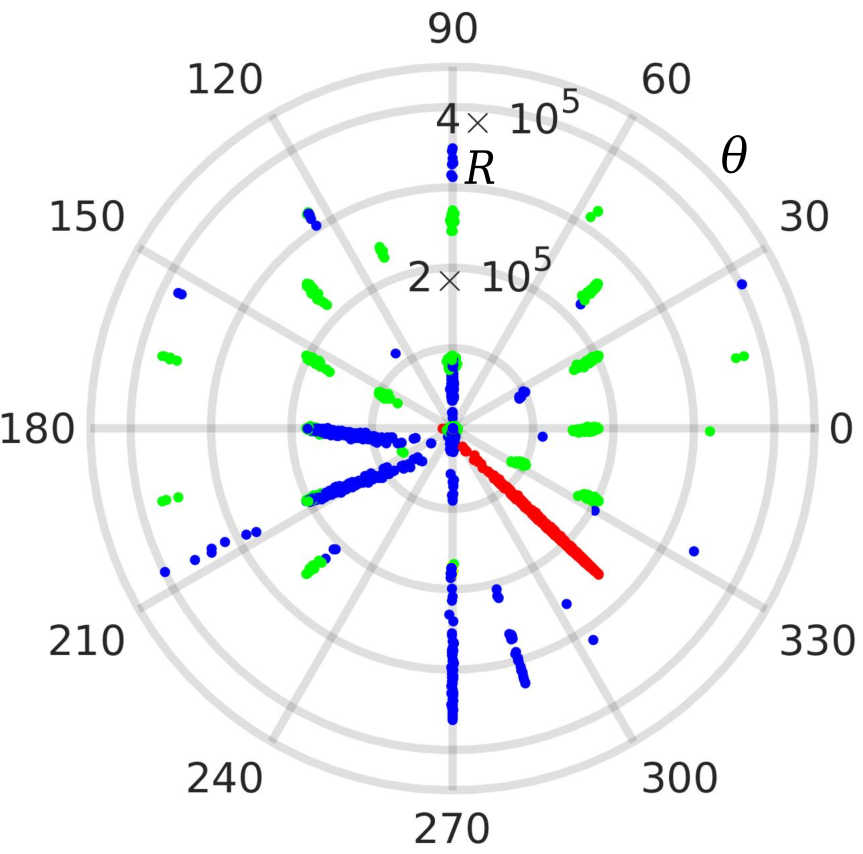}
\caption{Snapshot of particle positions at $t=5\times10^4 T$ (in the radial $R$ and angular $\theta$ coordinates) showing segregation of 3 different masses 0.23 (red), 0.87(green) and 1.33 (blue) corresponding to atoms Na, Rb and Cs with masses 23, 87 and 133 respectively commonly used in cold atom experiments. Remaining parameters are $\gamma=0.01$, $d_x=d_y=3$, $\omega=1$, $2k_x=k_y=2$ and $V=0.1679$. Since in our setup the dimensionless parameters depend on the ratios $\frac{V}{m}$, $\frac{d_{x,y}}{m}$ and$\frac{\gamma}{m}$, the parameters $V$, $d_{x,y}$ and $\gamma$ can be scaled up by the same factor as the atomic mass to correspond to relevant experimental setups.} \label{fig5}
\end{figure}

The segregation scheme mentioned here works not just for colloidal particles with a continuous size distribution but for example also for a mixture of 
particles differing only in mass. We now demonstrate this using a mixture of three cold thermal alkali atoms which can be treated classically in the regime of microkelvin temperatures \cite{Brown2008, Renzoni2009}. Such a scheme can be tested of course also with two species only and also applies to isotopic mixtures of the same atom (see \cite{Best2009, LeBlanc2007} for details on simultaeous trapping). In Fig.~\ref{fig5}, we show how our scheme can be adapted to simultaneously separate three species of commonly used alkali atoms of different masses from a mixture. It shows that after $t=5\times10^4 T$, particle rays containing mostly one species emerge at different angles. A detector positioned e.g. at $\theta \approx 320^{\circ}$ would collect only Na atoms whereas a detector at $\theta \approx 270^{\circ}$ would see mostly Cs atoms. Such a scheme may be realized using state of the art cold atom experimental setups with optical lattices driven by phase modulation of the laser beams using acousto-optical modulators and radio frequency generators \cite{Renzoni2009, Gommers2005, Lebedev2009, Eckardt2017}.

One advantage of our scheme compared to many other ratchet based segregation schemes is that is does not depend on noise and should be applicable 
to e.g. granular particles which are too heavy to allow Brownian noise to play a significant role and are often underdamped. 
However, as noise always accompanies dissipation, and may play a significant role for particles on colloidal scales, 
we have tested the robustness of the present scheme against white noise of strength typical for experiments corresponding to underdamped colloidal ratchets \cite{Hanggi2009} and cold atoms \cite{Cubero2010}. It adds only minor fluctuations around the average velocity of the limit cycle attractor which broadens the angular particle streams without essentially affecting the overall functionality of our segregation mechanism.

\paragraph{Conclusions} 
We have presented a scheme allowing to separate, or sort, particles 
from a mixture based on different selection criteria like radius-dependent frictional forces or particle mass. 
This scheme exploits the strong nonlinearity of driven lattices to control the late-time particle dynamics species selectively on the phase space level. This contrasts standard segregation schemes based on overdamped ratchet setups and allows us to overcome their key limitation of segregating more than two species. Owing to its deterministic character, our new control mechanism can be applied to particle mixtures on an unusually broad range of scales ranging from atoms to granular particles. The segregation scheme can be tested, for example, using polydisperse colloids or mixtures of cold thermal alkali atoms
using ac-driven optical lattices. As a perspective, further studies may account for localized perturbations of 
the ideal periodic potential employed which may allow to 
transfer the scheme even to the pure Hamiltonian regime based on e.g. a mass-selective accumulation 
of particles in the regular structures of the Hamiltonian phase space \cite{Wulf2014}. Interaction effects may add to the species-selective directed transport \cite{Liebchen2015}.

\paragraph{Acknowlegdements}
B.L. gratefully acknowledges funding  by  a  Marie  Curie  Intra  European  Fellowship (G.A.  no  654908)  within  Horizon  2020. A.K.M acknowledges a doctoral research grant (Funding ID: 57129429) by the Deutscher Akademischer Austauschdienst (DAAD). The authors thank T. Wulf for insightful discussions and C. Weitenberg for a careful reading of the manuscript. 

\bibliography{mybib}

\begin{thebibliography}{36}%
\makeatletter
\providecommand \@ifxundefined [1]{%
 \@ifx{#1\undefined}
}%
\providecommand \@ifnum [1]{%
 \ifnum #1\expandafter \@firstoftwo
 \else \expandafter \@secondoftwo
 \fi
}%
\providecommand \@ifx [1]{%
 \ifx #1\expandafter \@firstoftwo
 \else \expandafter \@secondoftwo
 \fi
}%
\providecommand \natexlab [1]{#1}%
\providecommand \enquote  [1]{``#1''}%
\providecommand \bibnamefont  [1]{#1}%
\providecommand \bibfnamefont [1]{#1}%
\providecommand \citenamefont [1]{#1}%
\providecommand \href@noop [0]{\@secondoftwo}%
\providecommand \href [0]{\begingroup \@sanitize@url \@href}%
\providecommand \@href[1]{\@@startlink{#1}\@@href}%
\providecommand \@@href[1]{\endgroup#1\@@endlink}%
\providecommand \@sanitize@url [0]{\catcode `\\12\catcode `\$12\catcode
  `\&12\catcode `\#12\catcode `\^12\catcode `\_12\catcode `\%12\relax}%
\providecommand \@@startlink[1]{}%
\providecommand \@@endlink[0]{}%
\providecommand \url  [0]{\begingroup\@sanitize@url \@url }%
\providecommand \@url [1]{\endgroup\@href {#1}{\urlprefix }}%
\providecommand \urlprefix  [0]{URL }%
\providecommand \Eprint [0]{\href }%
\providecommand \doibase [0]{http://dx.doi.org/}%
\providecommand \selectlanguage [0]{\@gobble}%
\providecommand \bibinfo  [0]{\@secondoftwo}%
\providecommand \bibfield  [0]{\@secondoftwo}%
\providecommand \translation [1]{[#1]}%
\providecommand \BibitemOpen [0]{}%
\providecommand \bibitemStop [0]{}%
\providecommand \bibitemNoStop [0]{.\EOS\space}%
\providecommand \EOS [0]{\spacefactor3000\relax}%
\providecommand \BibitemShut  [1]{\csname bibitem#1\endcsname}%
\let\auto@bib@innerbib\@empty
\bibitem [{\citenamefont {Jin}\ \emph {et~al.}(2014)\citenamefont {Jin},
  \citenamefont {McFaul}, \citenamefont {Duffy}, \citenamefont {Deng},
  \citenamefont {Tavassoli}, \citenamefont {Black},\ and\ \citenamefont
  {Ma}}]{Jin2014}%
  \BibitemOpen
  \bibfield  {author} {\bibinfo {author} {\bibfnamefont {C.}~\bibnamefont
  {Jin}}, \bibinfo {author} {\bibfnamefont {S.~M.}\ \bibnamefont {McFaul}},
  \bibinfo {author} {\bibfnamefont {S.~P.}\ \bibnamefont {Duffy}}, \bibinfo
  {author} {\bibfnamefont {X.}~\bibnamefont {Deng}}, \bibinfo {author}
  {\bibfnamefont {P.}~\bibnamefont {Tavassoli}}, \bibinfo {author}
  {\bibfnamefont {P.~C.}\ \bibnamefont {Black}}, \ and\ \bibinfo {author}
  {\bibfnamefont {H.}~\bibnamefont {Ma}},\ }\bibfield  {title} {\enquote
  {\bibinfo {title} {{Technologies for label-free separation of circulating
  tumor cells: from historical foundations to recent developments.}}}\ }\href
  {\doibase 10.1039/c3lc50625h} {\bibfield  {journal} {\bibinfo  {journal} {Lab
  Chip}\ }\textbf {\bibinfo {volume} {14}},\ \bibinfo {pages} {32} (\bibinfo
  {year} {2014})}\BibitemShut {NoStop}%
\bibitem [{\citenamefont {H{\"{a}}nggi}\ and\ \citenamefont
  {Marchesoni}(2009)}]{Hanggi2009}%
  \BibitemOpen
  \bibfield  {author} {\bibinfo {author} {\bibfnamefont {P.}~\bibnamefont
  {H{\"{a}}nggi}}\ and\ \bibinfo {author} {\bibfnamefont {F.}~\bibnamefont
  {Marchesoni}},\ }\bibfield  {title} {\enquote {\bibinfo {title} {{Artificial
  Brownian motors: Controlling transport on the nanoscale}},}\ }\href {\doibase
  10.1103/RevModPhys.81.387} {\bibfield  {journal} {\bibinfo  {journal} {Rev.
  Mod. Phys.}\ }\textbf {\bibinfo {volume} {81}},\ \bibinfo {pages} {387}
  (\bibinfo {year} {2009})}\BibitemShut {NoStop}%
\bibitem [{\citenamefont {Mullin}(2000)}]{Mullin2000}%
  \BibitemOpen
  \bibfield  {author} {\bibinfo {author} {\bibfnamefont {T.}~\bibnamefont
  {Mullin}},\ }\bibfield  {title} {\enquote {\bibinfo {title} {{Coarsening of
  self-organized clusters in binary mixtures of particles.}}}\ }\href {\doibase
  10.1103/PhysRevLett.84.4741} {\bibfield  {journal} {\bibinfo  {journal}
  {Phys. Rev. Lett.}\ }\textbf {\bibinfo {volume} {84}},\ \bibinfo {pages}
  {4741} (\bibinfo {year} {2000})}\BibitemShut {NoStop}%
\bibitem [{\citenamefont {Shinbrot}\ and\ \citenamefont
  {Muzzio}(1998)}]{Shinbrot1998}%
  \BibitemOpen
  \bibfield  {author} {\bibinfo {author} {\bibfnamefont {T.}~\bibnamefont
  {Shinbrot}}\ and\ \bibinfo {author} {\bibfnamefont {F.~J.}\ \bibnamefont
  {Muzzio}},\ }\bibfield  {title} {\enquote {\bibinfo {title} {{Reverse
  buoyancy in shaken granular beds}},}\ }\href {\doibase
  10.1103/PhysRevLett.81.4365} {\bibfield  {journal} {\bibinfo  {journal}
  {Phys. Rev. Lett.}\ }\textbf {\bibinfo {volume} {81}},\ \bibinfo {pages}
  {4365} (\bibinfo {year} {1998})}\BibitemShut {NoStop}%
\bibitem [{\citenamefont {Hong}\ \emph {et~al.}(2001)\citenamefont {Hong},
  \citenamefont {Quinn},\ and\ \citenamefont {Luding}}]{Hong2001}%
  \BibitemOpen
  \bibfield  {author} {\bibinfo {author} {\bibfnamefont {D.~C.}\ \bibnamefont
  {Hong}}, \bibinfo {author} {\bibfnamefont {P.~V.}\ \bibnamefont {Quinn}}, \
  and\ \bibinfo {author} {\bibfnamefont {S.}~\bibnamefont {Luding}},\
  }\bibfield  {title} {\enquote {\bibinfo {title} {{Reverse Brazil nut problem:
  Competition between percolation and condensation}},}\ }\href {\doibase
  10.1103/PhysRevLett.86.3423} {\bibfield  {journal} {\bibinfo  {journal}
  {Phys. Rev. Lett.}\ }\textbf {\bibinfo {volume} {86}},\ \bibinfo {pages}
  {3423} (\bibinfo {year} {2001})}\BibitemShut {NoStop}%
\bibitem [{\citenamefont {Bouzat}(2010)}]{Bouzat2010}%
  \BibitemOpen
  \bibfield  {author} {\bibinfo {author} {\bibfnamefont {S.}~\bibnamefont
  {Bouzat}},\ }\bibfield  {title} {\enquote {\bibinfo {title} {{Inertial
  effects, mass separation and rectification power in L\'{e}vy ratchets}},}\
  }\href {\doibase 10.1016/j.physa.2010.05.022} {\bibfield  {journal} {\bibinfo
   {journal} {Phys. A Stat. Mech. its Appl.}\ }\textbf {\bibinfo {volume}
  {389}},\ \bibinfo {pages} {3933} (\bibinfo {year} {2010})}\BibitemShut
  {NoStop}%
\bibitem [{\citenamefont {Zeng}\ \emph {et~al.}(2010)\citenamefont {Zeng},
  \citenamefont {Gong},\ and\ \citenamefont {Tian}}]{Zeng2010}%
  \BibitemOpen
  \bibfield  {author} {\bibinfo {author} {\bibfnamefont {C.}~\bibnamefont
  {Zeng}}, \bibinfo {author} {\bibfnamefont {A.}~\bibnamefont {Gong}}, \ and\
  \bibinfo {author} {\bibfnamefont {Y.}~\bibnamefont {Tian}},\ }\bibfield
  {title} {\enquote {\bibinfo {title} {{Current reversal and mass separation of
  inertial Brownian motors in a two-noise ratchet}},}\ }\href {\doibase
  10.1016/j.physa.2009.12.059} {\bibfield  {journal} {\bibinfo  {journal}
  {Phys. A Stat. Mech. its Appl.}\ }\textbf {\bibinfo {volume} {389}},\
  \bibinfo {pages} {1971} (\bibinfo {year} {2010})}\BibitemShut {NoStop}%
\bibitem [{\citenamefont {Marchesoni}(1998)}]{Marchesoni1998}%
  \BibitemOpen
  \bibfield  {author} {\bibinfo {author} {\bibfnamefont {F.}~\bibnamefont
  {Marchesoni}},\ }\bibfield  {title} {\enquote {\bibinfo {title} {{Conceptual
  design of a molecular shuttle}},}\ }\href {\doibase
  10.1016/S0375-9601(97)00841-4} {\bibfield  {journal} {\bibinfo  {journal}
  {Phys. Lett. A}\ }\textbf {\bibinfo {volume} {237}},\ \bibinfo {pages} {126}
  (\bibinfo {year} {1998})}\BibitemShut {NoStop}%
\bibitem [{\citenamefont {Astumian}\ and\ \citenamefont
  {H{\"{a}}nggi}(2002)}]{Astumian2002}%
  \BibitemOpen
  \bibfield  {author} {\bibinfo {author} {\bibfnamefont {R.~D.}\ \bibnamefont
  {Astumian}}\ and\ \bibinfo {author} {\bibfnamefont {P.}~\bibnamefont
  {H{\"{a}}nggi}},\ }\bibfield  {title} {\enquote {\bibinfo {title} {{Brownian
  Motors}},}\ }\href {\doibase 10.1063/1.1535005} {\bibfield  {journal}
  {\bibinfo  {journal} {Phys. Today}\ }\textbf {\bibinfo {volume} {55}},\
  \bibinfo {pages} {33} (\bibinfo {year} {2002})}\BibitemShut {NoStop}%
\bibitem [{\citenamefont {H{\"{a}}nggi}\ \emph {et~al.}(2005)\citenamefont
  {H{\"{a}}nggi}, \citenamefont {Marchesoni},\ and\ \citenamefont
  {Nori}}]{Hanggi2005}%
  \BibitemOpen
  \bibfield  {author} {\bibinfo {author} {\bibfnamefont {P.}~\bibnamefont
  {H{\"{a}}nggi}}, \bibinfo {author} {\bibfnamefont {F.}~\bibnamefont
  {Marchesoni}}, \ and\ \bibinfo {author} {\bibfnamefont {F.}~\bibnamefont
  {Nori}},\ }\bibfield  {title} {\enquote {\bibinfo {title} {{Brownian
  motors}},}\ }\href {\doibase 10.1002/andp.200410121} {\bibfield  {journal}
  {\bibinfo  {journal} {Ann. Phys.}\ }\textbf {\bibinfo {volume} {14}},\
  \bibinfo {pages} {51} (\bibinfo {year} {2005})}\BibitemShut {NoStop}%
\bibitem [{\citenamefont {Matthias}\ and\ \citenamefont
  {M{\"{u}}ller}(2003)}]{Matthias2003}%
  \BibitemOpen
  \bibfield  {author} {\bibinfo {author} {\bibfnamefont {S.}~\bibnamefont
  {Matthias}}\ and\ \bibinfo {author} {\bibfnamefont {F.}~\bibnamefont
  {M{\"{u}}ller}},\ }\bibfield  {title} {\enquote {\bibinfo {title}
  {{Asymmetric pores in a silicon membrane acting as massively parallel
  brownian ratchets}},}\ }\href {\doibase 10.1038/nature01736} {\bibfield
  {journal} {\bibinfo  {journal} {Nature (London)}\ }\textbf {\bibinfo {volume}
  {424}},\ \bibinfo {pages} {53} (\bibinfo {year} {2003})}\BibitemShut
  {NoStop}%
\bibitem [{\citenamefont {van Oudenaarden}(1999)}]{VanOudenaarden1999}%
  \BibitemOpen
  \bibfield  {author} {\bibinfo {author} {\bibfnamefont {A.}~\bibnamefont {van
  Oudenaarden}},\ }\bibfield  {title} {\enquote {\bibinfo {title} {{Brownian
  Ratchets: Molecular Separations in Lipid Bilayers Supported on Patterned
  Arrays}},}\ }\href {\doibase 10.1126/science.285.5430.1046} {\bibfield
  {journal} {\bibinfo  {journal} {Science}\ }\textbf {\bibinfo {volume}
  {285}},\ \bibinfo {pages} {1046} (\bibinfo {year} {1999})}\BibitemShut
  {NoStop}%
\bibitem [{\citenamefont {Liu}\ \emph {et~al.}(2016)\citenamefont {Liu},
  \citenamefont {Jiang}, \citenamefont {Tan}, \citenamefont {Yadav},
  \citenamefont {Biswas}, \citenamefont {van~der Maarel}, \citenamefont
  {Nijhuis},\ and\ \citenamefont {van Kan}}]{Liu2016}%
  \BibitemOpen
  \bibfield  {author} {\bibinfo {author} {\bibfnamefont {F.}~\bibnamefont
  {Liu}}, \bibinfo {author} {\bibfnamefont {L.}~\bibnamefont {Jiang}}, \bibinfo
  {author} {\bibfnamefont {H.~M.}\ \bibnamefont {Tan}}, \bibinfo {author}
  {\bibfnamefont {A.}~\bibnamefont {Yadav}}, \bibinfo {author} {\bibfnamefont
  {P.}~\bibnamefont {Biswas}}, \bibinfo {author} {\bibfnamefont {J.~R.~C.}\
  \bibnamefont {van~der Maarel}}, \bibinfo {author} {\bibfnamefont {C.~A.}\
  \bibnamefont {Nijhuis}}, \ and\ \bibinfo {author} {\bibfnamefont {J.~A.}\
  \bibnamefont {van Kan}},\ }\bibfield  {title} {\enquote {\bibinfo {title}
  {{Separation of superparamagnetic particles through ratcheted Brownian motion
  and periodically switching magnetic fields}},}\ }\href {\doibase
  10.1063/1.4967965} {\bibfield  {journal} {\bibinfo  {journal}
  {Biomicrofluidics}\ }\textbf {\bibinfo {volume} {10}},\ \bibinfo {pages}
  {064105} (\bibinfo {year} {2016})}\BibitemShut {NoStop}%
\bibitem [{\citenamefont {Reichhardt}\ and\ \citenamefont
  {Reichhardt}(2017)}]{Reichhardt2017}%
  \BibitemOpen
  \bibfield  {author} {\bibinfo {author} {\bibfnamefont {C.~J.~O.}\
  \bibnamefont {Reichhardt}}\ and\ \bibinfo {author} {\bibfnamefont
  {C.}~\bibnamefont {Reichhardt}},\ }\bibfield  {title} {\enquote {\bibinfo
  {title} {{Ratchet Effects in Active Matter Systems}},}\ }\href {\doibase
  10.1146/annurev-conmatphys-031016-025522} {\bibfield  {journal} {\bibinfo
  {journal} {Annu. Rev. Condens. Matter Phys.}\ }\textbf {\bibinfo {volume}
  {8}},\ \bibinfo {pages} {51} (\bibinfo {year} {2017})}\BibitemShut {NoStop}%
\bibitem [{\citenamefont {Ai}(2016)}]{Ai2016}%
  \BibitemOpen
  \bibfield  {author} {\bibinfo {author} {\bibfnamefont {B.}~\bibnamefont
  {Ai}},\ }\bibfield  {title} {\enquote {\bibinfo {title} {{Ratchet transport
  powered by chiral active particles}},}\ }\href {\doibase 10.1038/srep18740}
  {\bibfield  {journal} {\bibinfo  {journal} {Sci. Rep.}\ }\textbf {\bibinfo
  {volume} {6}},\ \bibinfo {pages} {18740} (\bibinfo {year}
  {2016})}\BibitemShut {NoStop}%
\bibitem [{\citenamefont {McFaul}\ \emph {et~al.}(2012)\citenamefont {McFaul},
  \citenamefont {Lin},\ and\ \citenamefont {Ma}}]{McFaul2012}%
  \BibitemOpen
  \bibfield  {author} {\bibinfo {author} {\bibfnamefont {S.~M.}\ \bibnamefont
  {McFaul}}, \bibinfo {author} {\bibfnamefont {B.~K.}\ \bibnamefont {Lin}}, \
  and\ \bibinfo {author} {\bibfnamefont {H.}~\bibnamefont {Ma}},\ }\bibfield
  {title} {\enquote {\bibinfo {title} {{Cell separation based on size and
  deformability using microfluidic funnel ratchets}},}\ }\href {\doibase
  10.1039/c2lc21045b} {\bibfield  {journal} {\bibinfo  {journal} {Lab Chip}\
  }\textbf {\bibinfo {volume} {12}},\ \bibinfo {pages} {2369} (\bibinfo {year}
  {2012})}\BibitemShut {NoStop}%
\bibitem [{\citenamefont {Petri}\ \emph {et~al.}(2011)\citenamefont {Petri},
  \citenamefont {Lenz}, \citenamefont {Liebchen}, \citenamefont {Diakonos},\
  and\ \citenamefont {Schmelcher}}]{Petri2011a}%
  \BibitemOpen
  \bibfield  {author} {\bibinfo {author} {\bibfnamefont {C.}~\bibnamefont
  {Petri}}, \bibinfo {author} {\bibfnamefont {F.}~\bibnamefont {Lenz}},
  \bibinfo {author} {\bibfnamefont {B.}~\bibnamefont {Liebchen}}, \bibinfo
  {author} {\bibfnamefont {F.}~\bibnamefont {Diakonos}}, \ and\ \bibinfo
  {author} {\bibfnamefont {P.}~\bibnamefont {Schmelcher}},\ }\bibfield  {title}
  {\enquote {\bibinfo {title} {{Formation of density waves via interface
  conversion of ballistic and diffusive motion}},}\ }\href {\doibase
  10.1209/0295-5075/95/30005} {\bibfield  {journal} {\bibinfo  {journal}
  {Europhys. Lett.}\ }\textbf {\bibinfo {volume} {95}},\ \bibinfo {pages}
  {30005} (\bibinfo {year} {2011})}\BibitemShut {NoStop}%
\bibitem [{\citenamefont {Wulf}\ \emph {et~al.}(2012)\citenamefont {Wulf},
  \citenamefont {Petri}, \citenamefont {Liebchen},\ and\ \citenamefont
  {Schmelcher}}]{Wulf2012}%
  \BibitemOpen
  \bibfield  {author} {\bibinfo {author} {\bibfnamefont {T.}~\bibnamefont
  {Wulf}}, \bibinfo {author} {\bibfnamefont {C.}~\bibnamefont {Petri}},
  \bibinfo {author} {\bibfnamefont {B.}~\bibnamefont {Liebchen}}, \ and\
  \bibinfo {author} {\bibfnamefont {P.}~\bibnamefont {Schmelcher}},\ }\bibfield
   {title} {\enquote {\bibinfo {title} {{Analysis of interface conversion
  processes of ballistic and diffusive motion in driven superlattices}},}\
  }\href {\doibase 10.1103/PhysRevE.86.016201} {\bibfield  {journal} {\bibinfo
  {journal} {Phys. Rev. E}\ }\textbf {\bibinfo {volume} {86}},\ \bibinfo
  {pages} {016201} (\bibinfo {year} {2012})}\BibitemShut {NoStop}%
\bibitem [{\citenamefont {Renzoni}(2009)}]{Renzoni2009}%
  \BibitemOpen
  \bibfield  {author} {\bibinfo {author} {\bibfnamefont {F.}~\bibnamefont
  {Renzoni}},\ }\bibfield  {title} {\enquote {\bibinfo {title} {{Driven
  Ratchets for Cold Atoms}},}\ }in\ \href {\doibase
  10.1016/S1049-250X(09)57001-2} {\emph {\bibinfo {booktitle} {Advances in
  Atomic Molecular and Optical Physics}}},\ Vol.~\bibinfo {volume} {57}\
  (\bibinfo  {publisher} {Academic Press},\ \bibinfo {year} {2009})\BibitemShut
  {NoStop}%
\bibitem [{\citenamefont {Reimann}(2002)}]{Reimann2002}%
  \BibitemOpen
  \bibfield  {author} {\bibinfo {author} {\bibfnamefont {P.}~\bibnamefont
  {Reimann}},\ }\bibfield  {title} {\enquote {\bibinfo {title} {{Brownian
  motors: noisy transport far from equilibrium}},}\ }\href {\doibase
  10.1016/S0370-1573(01)00081-3} {\bibfield  {journal} {\bibinfo  {journal}
  {Phys. Rep.}\ }\textbf {\bibinfo {volume} {361}},\ \bibinfo {pages} {57}
  (\bibinfo {year} {2002})}\BibitemShut {NoStop}%
\bibitem [{\citenamefont {Liebchen}\ \emph {et~al.}(2012)\citenamefont
  {Liebchen}, \citenamefont {Diakonos},\ and\ \citenamefont
  {Schmelcher}}]{Liebchen2012}%
  \BibitemOpen
  \bibfield  {author} {\bibinfo {author} {\bibfnamefont {B.}~\bibnamefont
  {Liebchen}}, \bibinfo {author} {\bibfnamefont {F.~K.}\ \bibnamefont
  {Diakonos}}, \ and\ \bibinfo {author} {\bibfnamefont {P.}~\bibnamefont
  {Schmelcher}},\ }\bibfield  {title} {\enquote {\bibinfo {title}
  {{Interaction-induced current-reversals in driven lattices}},}\ }\href
  {\doibase 10.1088/1367-2630/14/10/103032} {\bibfield  {journal} {\bibinfo
  {journal} {New J. Phys.}\ }\textbf {\bibinfo {volume} {14}},\ \bibinfo
  {pages} {103032} (\bibinfo {year} {2012})}\BibitemShut {NoStop}%
\bibitem [{\citenamefont {Mukhopadhyay}\ \emph {et~al.}(2016)\citenamefont
  {Mukhopadhyay}, \citenamefont {Liebchen}, \citenamefont {Wulf},\ and\
  \citenamefont {Schmelcher}}]{Mukhopadhyay2016}%
  \BibitemOpen
  \bibfield  {author} {\bibinfo {author} {\bibfnamefont {A.~K.}\ \bibnamefont
  {Mukhopadhyay}}, \bibinfo {author} {\bibfnamefont {B.}~\bibnamefont
  {Liebchen}}, \bibinfo {author} {\bibfnamefont {T.}~\bibnamefont {Wulf}}, \
  and\ \bibinfo {author} {\bibfnamefont {P.}~\bibnamefont {Schmelcher}},\
  }\bibfield  {title} {\enquote {\bibinfo {title} {{Freezing, accelerating, and
  slowing directed currents in real time with superimposed driven lattices}},}\
  }\href {\doibase 10.1103/PhysRevE.93.052219} {\bibfield  {journal} {\bibinfo
  {journal} {Phys. Rev. E}\ }\textbf {\bibinfo {volume} {93}},\ \bibinfo
  {pages} {052219} (\bibinfo {year} {2016})}\BibitemShut {NoStop}%
\bibitem [{\citenamefont {Flach}\ \emph {et~al.}(2000)\citenamefont {Flach},
  \citenamefont {Yevtushenko},\ and\ \citenamefont {Zolotaryuk}}]{Flach2000}%
  \BibitemOpen
  \bibfield  {author} {\bibinfo {author} {\bibfnamefont {S.}~\bibnamefont
  {Flach}}, \bibinfo {author} {\bibfnamefont {O.}~\bibnamefont {Yevtushenko}},
  \ and\ \bibinfo {author} {\bibfnamefont {Y.}~\bibnamefont {Zolotaryuk}},\
  }\bibfield  {title} {\enquote {\bibinfo {title} {{Directed current due to
  broken time-space symmetry}},}\ }\href {\doibase 10.1103/PhysRevLett.84.2358}
  {\bibfield  {journal} {\bibinfo  {journal} {Phys. Rev. Lett.}\ }\textbf
  {\bibinfo {volume} {84}},\ \bibinfo {pages} {2358} (\bibinfo {year}
  {2000})}\BibitemShut {NoStop}%
\bibitem [{\citenamefont {Struck}\ \emph {et~al.}(2013)\citenamefont {Struck},
  \citenamefont {Weinberg}, \citenamefont {{\"{O}}lschl{\"{a}}ger},
  \citenamefont {Windpassinger}, \citenamefont {Simonet}, \citenamefont
  {Sengstock}, \citenamefont {H{\"{o}}ppner}, \citenamefont {Hauke},
  \citenamefont {Eckardt}, \citenamefont {Lewenstein},\ and\ \citenamefont
  {Mathey}}]{Struck2013}%
  \BibitemOpen
  \bibfield  {author} {\bibinfo {author} {\bibfnamefont {J.}~\bibnamefont
  {Struck}}, \bibinfo {author} {\bibfnamefont {M.}~\bibnamefont {Weinberg}},
  \bibinfo {author} {\bibfnamefont {C.}~\bibnamefont {{\"{O}}lschl{\"{a}}ger}},
  \bibinfo {author} {\bibfnamefont {P.}~\bibnamefont {Windpassinger}}, \bibinfo
  {author} {\bibfnamefont {J.}~\bibnamefont {Simonet}}, \bibinfo {author}
  {\bibfnamefont {K.}~\bibnamefont {Sengstock}}, \bibinfo {author}
  {\bibfnamefont {R.}~\bibnamefont {H{\"{o}}ppner}}, \bibinfo {author}
  {\bibfnamefont {P.}~\bibnamefont {Hauke}}, \bibinfo {author} {\bibfnamefont
  {A.}~\bibnamefont {Eckardt}}, \bibinfo {author} {\bibfnamefont
  {M.}~\bibnamefont {Lewenstein}}, \ and\ \bibinfo {author} {\bibfnamefont
  {L.}~\bibnamefont {Mathey}},\ }\bibfield  {title} {\enquote {\bibinfo {title}
  {{Engineering Ising-XY spin-models in a triangular lattice using tunable
  artificial gauge fields}},}\ }\href {\doibase 10.1038/nphys2750} {\bibfield
  {journal} {\bibinfo  {journal} {Nat. Phys.}\ }\textbf {\bibinfo {volume}
  {9}},\ \bibinfo {pages} {738} (\bibinfo {year} {2013})}\BibitemShut {NoStop}%
\bibitem [{\citenamefont {Grynberg}\ and\ \citenamefont
  {Robilliard}(2001)}]{Grynberg2001}%
  \BibitemOpen
  \bibfield  {author} {\bibinfo {author} {\bibfnamefont {G.}~\bibnamefont
  {Grynberg}}\ and\ \bibinfo {author} {\bibfnamefont {C.}~\bibnamefont
  {Robilliard}},\ }\bibfield  {title} {\enquote {\bibinfo {title} {{Cold atoms
  in dissipative optical lattices}},}\ }\href {\doibase
  10.1016/s0370-1573(01)00017-5} {\bibfield  {journal} {\bibinfo  {journal}
  {Phys. Rep.}\ }\textbf {\bibinfo {volume} {355}},\ \bibinfo {pages} {335}
  (\bibinfo {year} {2001})}\BibitemShut {NoStop}%
\bibitem [{\citenamefont {Garbow}\ \emph {et~al.}(1997)\citenamefont {Garbow},
  \citenamefont {Muller}, \citenamefont {Sch{\"{a}}tzel},\ and\ \citenamefont
  {Palberg}}]{Garbow1997}%
  \BibitemOpen
  \bibfield  {author} {\bibinfo {author} {\bibfnamefont {N.}~\bibnamefont
  {Garbow}}, \bibinfo {author} {\bibfnamefont {J.}~\bibnamefont {Muller}},
  \bibinfo {author} {\bibfnamefont {K.}~\bibnamefont {Sch{\"{a}}tzel}}, \ and\
  \bibinfo {author} {\bibfnamefont {T.}~\bibnamefont {Palberg}},\ }\bibfield
  {title} {\enquote {\bibinfo {title} {{High-resolution particle sizing by
  optical tracking of single colloidal particles}},}\ }\href {\doibase
  10.1016/S0378-4371(96)00349-4} {\bibfield  {journal} {\bibinfo  {journal}
  {Physica A}\ }\textbf {\bibinfo {volume} {235}},\ \bibinfo {pages} {291}
  (\bibinfo {year} {1997})}\BibitemShut {NoStop}%
\bibitem [{\citenamefont {Lee}\ \emph {et~al.}(2007)\citenamefont {Lee},
  \citenamefont {Roichman}, \citenamefont {Yi}, \citenamefont {Kim},
  \citenamefont {Yang}, \citenamefont {van Blaaderen}, \citenamefont {van
  Oostrum},\ and\ \citenamefont {Grier}}]{Lee2007}%
  \BibitemOpen
  \bibfield  {author} {\bibinfo {author} {\bibfnamefont {S.}~\bibnamefont
  {Lee}}, \bibinfo {author} {\bibfnamefont {Y.}~\bibnamefont {Roichman}},
  \bibinfo {author} {\bibfnamefont {G.}~\bibnamefont {Yi}}, \bibinfo {author}
  {\bibfnamefont {S.}~\bibnamefont {Kim}}, \bibinfo {author} {\bibfnamefont
  {S.}~\bibnamefont {Yang}}, \bibinfo {author} {\bibfnamefont {A.}~\bibnamefont
  {van Blaaderen}}, \bibinfo {author} {\bibfnamefont {P.}~\bibnamefont {van
  Oostrum}}, \ and\ \bibinfo {author} {\bibfnamefont {D.~G.}\ \bibnamefont
  {Grier}},\ }\bibfield  {title} {\enquote {\bibinfo {title} {{Characterizing
  and tracking single colloidal particles with video holographic
  microscopy}},}\ }\href {\doibase 10.1364/OE.15.018275} {\bibfield  {journal}
  {\bibinfo  {journal} {Opt. Express}\ }\textbf {\bibinfo {volume} {15}},\
  \bibinfo {pages} {18275} (\bibinfo {year} {2007})}\BibitemShut {NoStop}%
\bibitem [{\citenamefont {Brown}\ and\ \citenamefont
  {Renzoni}(2008)}]{Brown2008}%
  \BibitemOpen
  \bibfield  {author} {\bibinfo {author} {\bibfnamefont {M.}~\bibnamefont
  {Brown}}\ and\ \bibinfo {author} {\bibfnamefont {F.}~\bibnamefont
  {Renzoni}},\ }\bibfield  {title} {\enquote {\bibinfo {title} {{Ratchet effect
  in an optical lattice with biharmonic driving: A numerical analysis}},}\
  }\href {\doibase 10.1103/PhysRevA.77.033405} {\bibfield  {journal} {\bibinfo
  {journal} {Phys. Rev. A - At. Mol. Opt. Phys.}\ }\textbf {\bibinfo {volume}
  {77}},\ \bibinfo {pages} {033405} (\bibinfo {year} {2008})}\BibitemShut
  {NoStop}%
\bibitem [{\citenamefont {Best}\ \emph {et~al.}(2009)\citenamefont {Best},
  \citenamefont {Will}, \citenamefont {Schneider}, \citenamefont
  {Hackerm\"{u}ller}, \citenamefont {van Oosten}, \citenamefont {Bloch},\ and\
  \citenamefont {L\"{u}hmann}}]{Best2009}%
  \BibitemOpen
  \bibfield  {author} {\bibinfo {author} {\bibfnamefont {T.}~\bibnamefont
  {Best}}, \bibinfo {author} {\bibfnamefont {S.}~\bibnamefont {Will}}, \bibinfo
  {author} {\bibfnamefont {U.}~\bibnamefont {Schneider}}, \bibinfo {author}
  {\bibfnamefont {L.}~\bibnamefont {Hackerm\"{u}ller}}, \bibinfo {author}
  {\bibfnamefont {D.}~\bibnamefont {van Oosten}}, \bibinfo {author}
  {\bibfnamefont {I.}~\bibnamefont {Bloch}}, \ and\ \bibinfo {author}
  {\bibfnamefont {D.~S.}\ \bibnamefont {L\"{u}hmann}},\ }\bibfield  {title}
  {\enquote {\bibinfo {title} {{Role of interactions in Rb87-K40 Bose-Fermi
  mixtures in a 3D optical lattice}},}\ }\href {\doibase
  10.1103/PhysRevLett.102.030408} {\bibfield  {journal} {\bibinfo  {journal}
  {Phys. Rev. Lett.}\ }\textbf {\bibinfo {volume} {102}},\ \bibinfo {pages}
  {030408} (\bibinfo {year} {2009})}\BibitemShut {NoStop}%
\bibitem [{\citenamefont {LeBlanc}\ and\ \citenamefont
  {Thywissen}(2007)}]{LeBlanc2007}%
  \BibitemOpen
  \bibfield  {author} {\bibinfo {author} {\bibfnamefont {L.~J.}\ \bibnamefont
  {LeBlanc}}\ and\ \bibinfo {author} {\bibfnamefont {J.~H.}\ \bibnamefont
  {Thywissen}},\ }\bibfield  {title} {\enquote {\bibinfo {title}
  {{Species-specific optical lattices}},}\ }\href {\doibase
  10.1103/PhysRevA.75.053612} {\bibfield  {journal} {\bibinfo  {journal} {Phys.
  Rev. A}\ }\textbf {\bibinfo {volume} {75}},\ \bibinfo {pages} {053612}
  (\bibinfo {year} {2007})}\BibitemShut {NoStop}%
\bibitem [{\citenamefont {Gommers}\ \emph {et~al.}(2005)\citenamefont
  {Gommers}, \citenamefont {Bergamini},\ and\ \citenamefont
  {Renzoni}}]{Gommers2005}%
  \BibitemOpen
  \bibfield  {author} {\bibinfo {author} {\bibfnamefont {R.}~\bibnamefont
  {Gommers}}, \bibinfo {author} {\bibfnamefont {S.}~\bibnamefont {Bergamini}},
  \ and\ \bibinfo {author} {\bibfnamefont {F.}~\bibnamefont {Renzoni}},\
  }\bibfield  {title} {\enquote {\bibinfo {title} {{Dissipation-Induced
  Symmetry Breaking in a Driven Optical Lattice}},}\ }\href {\doibase
  10.1103/PhysRevLett.95.073003} {\bibfield  {journal} {\bibinfo  {journal}
  {Phys. Rev. Lett.}\ }\textbf {\bibinfo {volume} {95}},\ \bibinfo {pages}
  {073003} (\bibinfo {year} {2005})}\BibitemShut {NoStop}%
\bibitem [{\citenamefont {Lebedev}\ and\ \citenamefont
  {Renzoni}(2009)}]{Lebedev2009}%
  \BibitemOpen
  \bibfield  {author} {\bibinfo {author} {\bibfnamefont {V.}~\bibnamefont
  {Lebedev}}\ and\ \bibinfo {author} {\bibfnamefont {F.}~\bibnamefont
  {Renzoni}},\ }\bibfield  {title} {\enquote {\bibinfo {title}
  {{Two-dimensional rocking ratchet for cold atoms}},}\ }\href {\doibase
  10.1103/PhysRevA.80.023422} {\bibfield  {journal} {\bibinfo  {journal} {Phys.
  Rev. A}\ }\textbf {\bibinfo {volume} {80}},\ \bibinfo {pages} {023422}
  (\bibinfo {year} {2009})}\BibitemShut {NoStop}%
\bibitem [{\citenamefont {Eckardt}(2017)}]{Eckardt2017}%
  \BibitemOpen
  \bibfield  {author} {\bibinfo {author} {\bibfnamefont {A.}~\bibnamefont
  {Eckardt}},\ }\bibfield  {title} {\enquote {\bibinfo {title} {{Colloquium:
  Atomic quantum gases in periodically driven optical lattices}},}\ }\href
  {\doibase 10.1103/RevModPhys.89.011004} {\bibfield  {journal} {\bibinfo
  {journal} {Rev. Mod. Phys.}\ }\textbf {\bibinfo {volume} {89}},\ \bibinfo
  {pages} {011004} (\bibinfo {year} {2017})}\BibitemShut {NoStop}%
\bibitem [{\citenamefont {Cubero}\ \emph {et~al.}(2010)\citenamefont {Cubero},
  \citenamefont {Lebedev},\ and\ \citenamefont {Renzoni}}]{Cubero2010}%
  \BibitemOpen
  \bibfield  {author} {\bibinfo {author} {\bibfnamefont {D.}~\bibnamefont
  {Cubero}}, \bibinfo {author} {\bibfnamefont {V.}~\bibnamefont {Lebedev}}, \
  and\ \bibinfo {author} {\bibfnamefont {F.}~\bibnamefont {Renzoni}},\
  }\bibfield  {title} {\enquote {\bibinfo {title} {{Current reversals in a
  rocking ratchet: Dynamical versus symmetry-breaking mechanisms}},}\ }\href
  {\doibase 10.1103/PhysRevE.82.041116} {\bibfield  {journal} {\bibinfo
  {journal} {Phys. Rev. E}\ }\textbf {\bibinfo {volume} {82}},\ \bibinfo
  {pages} {041116} (\bibinfo {year} {2010})}\BibitemShut {NoStop}%
\bibitem [{\citenamefont {Wulf}\ \emph {et~al.}(2014)\citenamefont {Wulf},
  \citenamefont {Liebchen},\ and\ \citenamefont {Schmelcher}}]{Wulf2014}%
  \BibitemOpen
  \bibfield  {author} {\bibinfo {author} {\bibfnamefont {T.}~\bibnamefont
  {Wulf}}, \bibinfo {author} {\bibfnamefont {B.}~\bibnamefont {Liebchen}}, \
  and\ \bibinfo {author} {\bibfnamefont {P.}~\bibnamefont {Schmelcher}},\
  }\bibfield  {title} {\enquote {\bibinfo {title} {{Disorder Induced Regular
  Dynamics in Oscillating Lattices}},}\ }\href {\doibase
  10.1103/PhysRevLett.112.034101} {\bibfield  {journal} {\bibinfo  {journal}
  {Phys. Rev. Lett.}\ }\textbf {\bibinfo {volume} {112}},\ \bibinfo {pages}
  {034101} (\bibinfo {year} {2014})}\BibitemShut {NoStop}%
\bibitem [{\citenamefont {Liebchen}\ and\ \citenamefont
  {Schmelcher}(2015)}]{Liebchen2015}%
  \BibitemOpen
  \bibfield  {author} {\bibinfo {author} {\bibfnamefont {B.}~\bibnamefont
  {Liebchen}}\ and\ \bibinfo {author} {\bibfnamefont {P.}~\bibnamefont
  {Schmelcher}},\ }\bibfield  {title} {\enquote {\bibinfo {title} {{Interaction
  induced directed transport in ac-driven periodic potentials}},}\ }\href
  {\doibase 10.1088/1367-2630/17/8/083011} {\bibfield  {journal} {\bibinfo
  {journal} {New J. Phys.}\ }\textbf {\bibinfo {volume} {17}},\ \bibinfo
  {pages} {083011} (\bibinfo {year} {2015})}\BibitemShut {NoStop}%
\end{thebibliography}%

\end{document}